\documentclass[aps,preprint,nofootinbib,amsmath,amssymb,ascmac,12pt]{revtex4}
\usepackage{bm}
\usepackage[dvipdfm]{graphicx}
\usepackage{color}

\newcommand{\QM}{Q_{\rm M}}
\newcommand{\QE}{Q_{\rm E}}

\newcommand{\rg}{r_{\rm g}}

\newcommand{\Lx}{L_{(x)}}
\newcommand{\Ly}{L_{(y)}}
\newcommand{\Lz}{L_{(z)}}

\newcommand{\calT}{{\cal T}}
\newcommand{\calE}{{\cal E}}

\newcommand{\calK}{{\cal K}}
\newcommand{\calLx}{{\cal L}_{(x)}}
\newcommand{\calLy}{{\cal L}_{(y)}}
\newcommand{\calLz}{{\cal L}_{(z)}}
\newcommand{\calV}{{\cal V}}

\newcommand{\calQM}{{\cal Q}_{\rm M}}
\newcommand{\calQE}{{\cal Q}_{\rm E}}

\newcommand{\mel}{m_{\rm e}}
\newcommand{\mpr}{m_{\rm p}}

\newcommand{\epzero}{\varepsilon_0}
\newcommand{\kB}{k_{\rm B}}

\newcommand{\Tp}{T_{\rm p}}
\newcommand{\Te}{T_{\rm e}}

\newcommand{\Rmin}{R_{\rm min}}
\newcommand{\Rmax}{R_{\rm max}}

\begin{document}

\title{Motion of a charged test particle around a static black hole in a monopole magnetic field }

\author{Ken-ichi Nakao$^{1,2,3}$\footnote{E-mail:knakao@omu.ac.jp}}
\author{Yota Endo$^1$}
\author{Hideki Ishihara$^{2,3}$}
\author{Kenta Matsuo$^1$}
\author{Kensuke Sueto$^1$}
\author{Koudai Ueda$^{3}$}
\author{Hirotaka Yoshino$^{1,2,3}$}
\affiliation{
${}^{1}$
Department of Physics, Graduate School of Science, Osaka Metropolitan University, 3-3-138 Sugimoto, Sumiyoshi, Osaka 558-8585, Japan\\
${}^{2}$Nambu Yoichiro Institute for Theoretical and Experimental Physics (NITEP), Osaka Metropolitan University, 
3-3-138 Sugimoto, Sumiyoshi, Osaka 558-8585, Japan\\
${}^{3}$Osaka Central Advanced Mathematical Institute (OCAMI), 
Osaka Metropolitan University, 3-3-138 Sugimoto, Sumiyoshi, Osaka 558-8585, Japan}


\begin{abstract}
We study the motion of a charged test particle in the spacetime with a spherically symmetric black hole which is immersed in a monopole magnetic field. We show that the radial motion of the charged test particle is governed by completely the same equation as that in the case of no magnetic field. This result implies that the black hole will acquire the electric charge if it is surrounded by the collisionless plasma composed of protons and electrons which obey the Maxwell velocity distribution. The drastically different situation from no magnetic field case appears in the angular motions of  charged test particles due to the magnetic field. The trajectory of a charged test particle 
around the black hole in the magnetic field of the order of 10 Gauss near the 
event horizon is confined on a very thin cone as long as the specific angular momentum of the particle is not much larger than the gravitational radius of the black hole times the speed of light. This result leads to a possibility that a plasma lump can hover over the black hole and is very hot,  in the monopole magnetic field.  
\end{abstract}

\preprint{OCU-PHYS-618}
\preprint{AP-GR-206}
\preprint{NITEP 268}
\date{\today}
\maketitle

\section{Introduction}\label{sec:Intro}

A black hole is defined as a complement of the causal past of the future null infinity\cite{Hawking}, and 
is not observable by its definition\cite{Nakao}, whereas general relativity predicts that black holes exist in our universe. 
Black holes are predicted to be formed by the gravitational collapse of massive objects, and their mass 
may increase by absorbing surrounding matter. However, all we can observe is the black hole candidate 
that will form the black hole after infinite amount of time. 
Observable prediction related to the black hole, which is made by general relativity, is 
that the spacetime geometry around a stationary black hole is 
described by the Kerr-Newman family. A related issue is the black hole 
shadow\cite{Luminet}. The Event Horizon Telescope (EHT) Collaboration reported that 
the black hole shadows at the center of the M87 galaxy\cite{EHT-M87-1} and the Milky Way galaxy\cite{EHT-MW-1} were photographed, 
but the debate over the reports by EHT Collaboration remain unresolved\cite{MKM-1,MKM-2,MKM-3}.

Active Galactic Nuclei (AGN) are considered to be deeply related to black holes. 
The AGN are phenomena in each of which a narrow region at the center of a galaxy emits strong electromagnetic 
radiation that is almost equal to the entire galactic emission. The source of the radiative energy 
is thought to be the gravitational energy released by mass accretion 
onto the supermassive black hole \cite{Brandford-etal,Lynden-Bell,Bardeen,Rees,Antonucci}. 
Very long relativistic jets of plasma from AGNs are observed and their energy sources 
are regarded as one of important issues of the black hole astrophysics. 
Long jets implies that they should be steady for a long time which is equal to 
or longer than the length of the jet divided by the speed of light. 
The Brandford-Znajek process\cite{BZ-process} in which the rotational energy 
of the black hole is released through the neighboring magnetic field  is regarded as 
a promising candidate of the energy source of the relativistic jet. 

When we study physical phenomena around a black hole, the black hole is usually assumed to be electrically neutral. 
Wald, however, pointed out the possibility that in the case that a rotating black hole is immersed in a uniform magnetic field, 
since the rotation of the black hole causes a difference of an electrostatic potential between the event horizon and infinity, 
the black hole can acquire the electric charge, $8\pi\epzero G\vec{J}\cdot\vec{B}/c^2$, where $c$, $\epzero$, 
$G$, $\vec{J}$ and $\vec{B}$ are the speed of light, 
the electric permittivity of vacuum, Newton's gravitational constant, the angular momentum of the black hole and the 
magnetic field, respectively\cite{Wald-solution}. 
Recently, King and Pringle stated that if the rotating black hole gets the electric charge of Wald's value, 
the Brandford-Znajek process cannot provide energy sources of the {\it steady} relativistic jets from AGN's\cite{King-Pringle}. 
Komissarov refuted their statementm and this issue is still an open question. 
Anyway, the electrification of black holes will be important in the consideration of the black hole astrophysics\cite{Komissarov}. 

Zaja\v{c}ek et al. showed that even in the case of a non-rotating black hole with no magnetic field, 
the black hole can acquire an electric charge less than about $2\pi\epzero GM\mpr/e$ or two times of it, 
where $\mpr$ and $e$ are the mass of a proton and 
the elementary charge, respectively\cite{Zajacek-etal}. Since the analyses of Zaja\v{c}ek et al 
are basically based on the non-relativistic analysis, 
four of the present authors, KN, KM, HY, HI, studied the electrification 
of a static black hole by investigating the difference of the probabilities of 
a proton and an electron, whose initial velocity obey the Maxwell distribution, falling into the black hole\cite{paper-I}. 
Hereafter we call Ref.~\cite{paper-I} NMYI paper. In NMYI paper, the maximal 
electric charge acquired by a static black hole is the same as that obtained 
by Zaja\v{c}ek et al, i.e., $2\pi\epzero GM\mpr/e$, 
and the radii of the Innermost Stable Circular Orbits (ISCO) of a proton 
and an electron can be very larger than that of a neutral test particle 
around a static charged black hole. 

By those reasons mentioned above, in order to understand the physical situation around a black hole, it is crucially important to understand the motion of charged particles around the black hole immersed in electromagnetic field.  
In this paper, we study the motion of a charged test particle around a static black hole immersed in a monopole 
magnetic field. By using results obtained through this study, we investigate a 
possibility of the electrification of the black hole, and also make a comment on the temperature 
of the collisionless plasma surrounding the black hole.   
Here we should mention that Khan and Chen studied a charged test particle in the spacetime with a rotating black hole which is immersed in 
a split magnetic monopole field \cite{Khan-Chen}. However, their interest is on the magnetic Penrose process, and different from 
our present purposes.

This paper is organized as follows. In Sec.~II, we briefly review the spherically symmetric black hole in monopole 
electromagnetic field. 
In Sec.~III, we study the motion of a charged test particle around the spherically symmetric black hole 
in the monopole electromagnetic field.  
In Sec.~IV, we show that a non-rotating black hole in a monopole magnetic field will 
acquire the positive electric charge as in the case without the magnetic field. 
In Sec.~V, consider the effects of magnetic field on the temperature of the plasma surrounding the black hole 
by using the result obtained in Secs.~II and III. 
Section~VI is devoted for summary and discussion. 
In Appendix A, the probability of the test particle initially moving outward falling into the black hole is estimated. 
In Appendix B, we estimate the electric current on the surface of a lump of charged test particles. 

In this paper, we adopt the abstract index notation in which the Latin indices, except for $t$ and $r$ which are 
assigned to the time and radial coordinates, represent a type of a tensor, 
whereas the Greek indices,  except for $\theta$ and $\varphi$ which are assigned to the coordinates on a sphere, 
represent the components of a tensor with respect to the coordinate bases, 
and follow the sign convention of the metric tensor in Ref.~\cite{Wald} and adopt the SI unit. 
The covariant derivative and the ordinary derivative are denoted by $\nabla_a$ and $\partial_a$, respectively. 
The speed of light and Newton's gravitational constant are denoted by $c$ and $G$, respectively.

\section{Static black hole in a monopole electromagnetic field}

In this paper, we consider the electrification of a static black hole immersed in a monopole magnetic field. 
If there is no matter around the black hole, the spacetime geometry is described by 
the dyonic Reissner-Nordstr\"{o}m (RN) solution whose infinitesimal world interval is given by
\begin{equation}
ds^2=g_{\alpha\beta}dx^\alpha dx^\beta=-c^2f(r) dt^2+\frac{dr^2}{f(r)}+r^2\left(d\theta^2+\sin^2\theta d\varphi^2\right), \label{metric}
\end{equation}
with
\begin{equation}
f(r)=1-\frac{2GM}{c^2r}+\frac{Q^2}{r^2} \label{RN-f}
\end{equation}
where $Q$ and $M$ are the charge parameter and gravitational mass in this system, respectively, 

In the case that the electrified black hole is in a monopole magnetic field, the four-vector potential is given by
\begin{equation}
A_\mu=-\frac{\QE}{4\pi\varepsilon_0 r} \delta^t_\mu
-\frac{\QM}{4\pi\varepsilon_0 c}\cos\theta\delta^\varphi_\mu, \label{monopole}
\end{equation}
where $\QE$ and $\QM$ are the electric charge and the magnetic charge, of the black hole, respectively. 

The charge parameter $Q$ of the RN solution is related to 
the electric charge $Q_{\rm E}$ and the magnetic charge $Q_{\rm M}$ in the manner, 
\begin{equation}
Q^2=\frac{G}{4\pi\varepsilon_0 c^4}\left(Q_{\rm E}^2+Q_{\rm M}^2\right).
\label{conversion}
\end{equation}

In this paper, we focus on situations similar to the neighborhood of, for example, Sgr~A*,
or the center of M87.  
The mass of Sgr~A* is about $4\times 10^6M_\odot$ and the magnetic field in its neighborhood is estimated to be 
about $10$G \cite{EHT-MW-5}, whereas the mass of the black hole candidate in M87 is equal to about 
$6.5\times10^9 M_\odot$ and the magnetic field in its neighborhood is 
estimated be also about $10$G \cite{EHT-M87-5}. The magnetic charge $\QM$ which generates 
the magnetic field $B^a$ of the value $10$G at $r=N\rg$ is estimated as follows, where $N$ is a free parameter larger than one, and
\begin{equation}
\rg:=\frac{2GM}{c^2}
\end{equation}
is the gravitational radius of the black hole. We have the following magnetic field for static observers: 
\begin{equation}
B^\alpha=\frac{1}{2}\sqrt{c^2f(r)}\varepsilon^{t r\mu\nu}F_{\mu\nu}\delta^\alpha_r
=\frac{\sqrt{f(r)}}{r^2\sin\theta}\left(\partial_\theta A_\varphi-\partial_\varphi A_\theta\right)\delta^\alpha_r
=\frac{\QM\sqrt{f(r)}}{4\pi\epzero c r^2}\delta^\alpha_r,
\end{equation}
where $\varepsilon^{abcd}$ is the skew tensor whose component is given 
as $\varepsilon^{0123}=1/\sqrt{-g}$ with $g$ being the determinant of the metric tensor. Hence, we obtain 
\begin{equation}
\frac{\QM}{4\pi\epzero c}=\frac{r^2B^r}{\sqrt{f(r)}} =\left(\frac{N^5}{N-1}\right)^{1\over2}~ \rg^2 B_N, \label{QB-relation}
\end{equation}
where
\begin{equation}
B_N:=B^r|_{r=N\rg}.
\label{BN-def}
\end{equation}
By using Eqs.~\eqref{conversion} and \eqref{QB-relation}, in the case of $\QE=0$, we have
\begin{align}
\frac{1}{\rg}\times\frac{Q^2}{r}&=\frac{G\QM^2}{4\pi \varepsilon_0 c^4\rg^2}\left(\frac{\rg}{r}\right)
=\frac{4\pi\epzero G\rg^2}{c^2}\left(\frac{N^5}{N-1}\right)\left(\frac{\rg}{r}\right) B_N^2
\nonumber \\
&=1.1\times10^{-23}\left(\frac{\rg}{r}\right)\left(\frac{N^5}{N-1}\right)
\left(\frac{B_N}{10^{-3}{\rm T}}\right)^2
\left(\frac{M}{4\times10^6M_\odot}\right)^2.
\end{align}
As will be mentioned later, the energy of the electric field generated by the electric charge 
acquired by the black hole is also very small. Hence the spacetime geometry is well approximated 
by the Schwarzschild solution 
whose infinitesimal world interval is given by Eq.~\eqref{metric} with not Eq.~\eqref{RN-f} but
\begin{equation}
f(r)=1-\frac{\rg}{r}. \label{Sch-f}
\end{equation}
Thus, hereafter, we assume the Schwarzschild metric.

\section{Motion of a charged test particle in monopole field}\label{sec:EOM}

In this section, we study the motion of a charged test particle in 
the non-rotating black hole spacetime with the four-vector potential Eq.~\eqref{monopole}. 
The world line of a test particle is given in the form $x^\alpha=x^\alpha(\tau)$, where $\tau$ is the proper time of the test particle, 
and then the components of the four-velocity $u^a$ of the test particle with respect to the coordinate basis is defined as
\begin{equation}
u^\alpha=\frac{dx^\alpha}{d\tau}.
\end{equation}
The four-velocity $u^a$ satisfies the normalization condition 
\begin{equation}
u^a u_a=-c^2. \label{n-condition}
\end{equation}
The equation of motion for a charged test particle with the mass $m$ and the electric charge $q$ is 
 \begin{equation}
 mu^b\nabla_b u_a=q F_{ab}u^b, \label{EOM}
 \end{equation}
 where $\nabla_b$ is the covariant derivative, 
 whereas $F_{ab}$ is the electromagnetic field strength tensor defined as
 \begin{equation}
 F_{ab}=\nabla_a A_b-\nabla_b A_a=\partial_a A_b-\partial_b A_a,
 \end{equation}
 where $\partial_b$ is the ordinary derivative.

\subsection{Conserved quantities}

 If a Killing vector, $\xi^a$, exists, we have
\begin{align}
u^b\nabla_b \left(m u_a\xi^a\right)&
=\left(m u^b\nabla_b u_a\right)\xi^a+m u^a u^b\nabla_b \xi_a \nonumber \\
&=q \left(\nabla_a A_b-\nabla_b A_a\right)\xi^a u^b 
\nonumber \\
&=q \left[\nabla_a\left( A_b+\nabla_b\chi\right)-\nabla_b \left(A_a+\nabla_a\chi\right)\right]\xi^a u^b\nonumber \\
&=q\left\{{\cal L}_\xi \left(A_b+\nabla_b\chi\right)-\nabla_b\left[\left(A_a+\nabla_a\chi\right)\xi^a\right]\right\}u^b,
\end{align}
where we have used Eq.~\eqref{EOM}  and the Killing equation  $\nabla_{(a}\xi_{b)}=0$ in the second equality. 
Thus we have
\begin{equation}
u^b\nabla_b\Bigl\{\bigl[m u_a+q\left(A_a+\nabla_a\chi\right)\bigr]\xi^a\Bigr\}
=qu^a{\cal L}_\xi\left( A_a+\nabla_a\chi\right).
\end{equation}
If ${\cal L}_\xi \left(A_a+\nabla_a\chi\right)=0$ holds, this equation implies that 
$\bigl[m u_a+q\left(A_a+\nabla_a\chi\right)\bigr]\xi^a$ is conserved.

In the spacetime whose metric is given by Eq.~\eqref{metric}, 
the time coordinate basis vector $(\partial/\partial t)^a$ is a Killing vector, and 
there is the SO$(3)$ isometry whose generators are three Killing vectors given below. 
By defining the Cartesian coordinates $(x,y,z)$ for three-dimensional space as
\begin{equation}
(x,y,z)=(r\sin\theta\cos\varphi,r\sin\theta\sin\varphi,r\cos\theta),
\end{equation}
the three Killing vectors are given as  
\begin{align}
{L}_{(x)}^a&=y\left(\frac{\partial}{\partial z}\right)^a-z\left(\frac{\partial}{\partial y}\right)^a
=-\sin\varphi\left(\frac{\partial}{\partial\theta}\right)^a-\cot\theta\cos\varphi\left(\frac{\partial}{\partial\varphi}\right)^a, \\
{L}_{(y)}^a&=z\left(\frac{\partial}{\partial x}\right)^a-x\left(\frac{\partial}{\partial z}\right)^a
=\cos\varphi\left(\frac{\partial}{\partial\theta}\right)^a-\cot\theta\sin\varphi\left(\frac{\partial}{\partial\varphi}\right)^a, \\
{L}_{(z)}^a&=x\left(\frac{\partial}{\partial y}\right)^a-y\left(\frac{\partial}{\partial x}\right)^a=\left(\frac{\partial}{\partial\varphi}\right)^a,
\end{align}
where $\left(\dfrac{\partial}{\partial x^\mu}\right)^a$ denotes a coordinate basis vector. 
These Killing vectors generate the rotations around $x$-axis, $y$-axis and $z$-axis, respectively.

We introduce two functions, $\chi_{(x)}$ and $\chi_{(y)}$, defined as solutions of the following differential equations. 
\begin{align}
\Lx^a\nabla_a \chi_{(x)}&=-\frac{\QM}{4\pi\epzero c}\frac{\cos\varphi}{\sin\theta}, \\
\Ly^a\nabla_a \chi_{(y)}&=-\frac{\QM}{4\pi\epzero c}\frac{\sin\varphi}{\sin\theta}.
\end{align}
The derivatives in these equations are the ordinary differentiations along the integral curves of the Killing vectors $L_{(x)}^a$ and 
$L_{(y)}^a$, respectively, and hence the solutions exist.  

Then the following equations hold.
\begin{align}
{\cal L}_{\partial\over\partial t} A_a&=0, \\
{\cal L}_{\Lx}\left(A_a+\nabla_a\chi_{(x)}\right)&=0, \\
{\cal L}_{\Ly}\left(A_a+\nabla_a\chi_{(y)}\right)&=0, \\
{\cal L}_{\Lz}A_a&=0,
\end{align}
where note that both of ${\cal L}_{\Lx}A_a$ and ${\cal L}_{\Ly}A_a$ do not vanish. 
Thus we have the four conserved quantities given as
\begin{align}
E&:=-\left(u_a+\frac{q}{m}A_a\right)\left(\frac{\partial}{\partial t}\right)^a \nonumber \\
&=-u_t+\frac{q\QE}{4\pi\epzero mc r}, \\
\Lx&:=\left[u_a+\frac{q}{m}\left(A_a+\nabla_a\chi_{(x)}\right)\right]L_{(x)}^a \nonumber \\
&~=-u_\theta\sin\varphi-u_\varphi\cot\theta\cos\varphi-\frac{q\QM}{4\pi\epzero mc} \sin\theta\cos\varphi, \label{Lx-def}\\
\Ly&:=\left[u_a+\frac{q}{m}\left(A_a+\nabla_a\chi_{(y)}\right)\right]L_{(y)}^a \nonumber \\
&~=u_\theta\cos\varphi-u_\varphi\cot\theta\sin\varphi-\frac{q\QM}{4\pi\epzero mc}\sin\theta\sin\varphi, \label{Ly-def}\\
\Lz&:=\left(u_a+\frac{q}{m}A_a\right)L_{(z)}^a \nonumber \\
&=u_\varphi-\frac{q\QM}{4\pi\epzero mc}\cos\theta \label{Lz-def}
\end{align}
where $E$, $\Lx$, $\Ly$ and $\Lz$ correspond to the specific energy, 
the $x$-, $y$- and $z$-components of the specific angular momentum, respectively. 

From Eqs.~\eqref{Lx-def}--\eqref{Lz-def}, we have
\begin{equation}
u_\theta^2+\frac{u_\varphi^2}{\sin^2\theta}=\Lx^2+\Ly^2+\Lz^2-\left(\frac{q\QM}{4\pi\epzero mc}\right)^2=:K^2. \label{K-def}
\end{equation}

\subsection{Motion in radial direction}

For notational simplicity, we introduce the dimensionless quantities defined as 
\begin{align}
R&:=\frac{r}{\rg}, \label{dimless-r}\\
\calT&:=\frac{c\tau}{\rg}, \label{dimless-tau}\\
\calE&:=\frac{E}{c}, \label{dimless-E}\\
\left(\calLx,\calLy,\calLz\right)&:=\left(\frac{\Lx}{c\rg},\frac{\Ly}{c\rg},\frac{\Lz}{c\rg}\right), \label{dimless-L} \\
\calK&:=\frac{K}{c\rg}, \label{ell-def} \\
\calQE&:=\frac{q\QE}{4\pi\epzero mc^2\rg}, \\
\calQM&:=\frac{q\QM}{4\pi\epzero mc^2\rg}, \label{calQM}
\end{align}

From the normalization of the four velocity, Eq.~\eqref{n-condition}, we have
\begin{equation}
\left(\frac{dR}{d\calT}\right)^2+U(R;\calE,\calK^2,\calQE)=0, \label{energy-eq}
\end{equation}
where
\begin{equation}
U(R;\calE,\calK^2,\calQE):=-\left(\calE-\frac{\calQE}{R}\right)^2+\left(1-\frac{1}{R}\right)\left(1+\frac{\calK^2}{R^2}\right). \label{potential}
\end{equation}
Equation \eqref{energy-eq} is identical to the corresponding equation for the charged test particle 
in the Schwarzschild spacetime with no magnetic charge, $\QM=0$, but with non-vanishing electric charge, $\QE\neq0$.  
No effect of the monopole magnetic 
field appears in the radial motion of the charged test particle, and hence, for example, the radius of ISCO   
is unchanged from that with no magnetic field.\footnote{It should be noted 
that ISCO radius the rotating black hole immersed in the asymptotically homogeneous magnetic field 
can be very different from the case without magnetic field \cite{IHK2012}. }

We write $U(R;\calE,\calK^2,\calQE)$ in the form
\begin{align}
U(R;\calE,\calK^2,\calQE)=-\left[
\calE-\frac{\calQE}{R}+\sqrt{\left(1-\frac{1}{R}\right)\left(1+\frac{\calK^2}{R^2}\right)}
\right]\left[{\cal E}-V(R;\calK^2,\calQE)\right]
\end{align}
where
\begin{equation}
V(R;\calK^2,\calQE)=\frac{\calQE}{R}+\sqrt{\left(1-\frac{1}{R}\right)\left(1+\frac{\calK^2}{R^2}\right)}. \label{V-def}
\end{equation}
By noting that 
\begin{equation}
f(r)\frac{dt}{d\tau}=\calE-\frac{\calQE}{R}>0~~~{\rm for}~r>\rg
\end{equation}
must hold by the future directed property of the four-velocity, we find that the allowed domain for the motion of the charged test particle is determined by the inequality
\begin{equation}
V(R;\calK^2,\calQE)\leq\calE.
\end{equation}

\subsection{Motion in angular directions}\label{A-motion}

Here, we adopt the coordinates in which the angular momentum of the charged test particle 
is parallel to the $z$-axis, i.e., $\Lx=\Ly=0$. We have, from Eqs.~\eqref{Lx-def} and \eqref{Ly-def}, 
\begin{equation}
u_\theta=-\Lx \sin\varphi+\Ly\cos\varphi,
\end{equation}
and hence $u_\theta=r^2 \dfrac{d\theta}{d\tau}=0$ holds in the coordinates we adopt. The motion of the 
charged test particle is restricted in a hypersurface of constant polar angle 
\begin{equation}
\theta=\vartheta.
\end{equation}  

Substituting $\Lx=\Ly=0$, $u_\theta=0$ and $\theta=\vartheta$ into Eqs.~\eqref{Lx-def} and \eqref{Ly-def}, we have
\begin{equation}
u_\varphi\cos\vartheta+\calQM c\rg \sin^2\vartheta=0. \label{u-ph}
\end{equation}
Substituting this equation into Eq.~\eqref{Lz-def} to eliminate $u_\varphi$, we have
\begin{equation}
\calLz\cos\vartheta+\calQM=0 \label{eq:C0}
\end{equation}
We find, from this equation, that the sign of $\calLz$ is opposite to that of $\calQM$ in $0\leq\vartheta<\pi/2$, 
whereas the sign of $\calLz$ is the same as that of $\calQM$ in $\pi/2 \leq\vartheta\leq\pi$. 
Eliminating $\calQM$ from Eq.~\eqref{u-ph} by using Eq.~\eqref{eq:C0}, we obtain
\begin{equation}
u_\varphi=\calLz c\rg\sin^2\vartheta. \label{u-Lz}
\end{equation}
This equation implies that the sign of $\dfrac{d\varphi}{d\tau}$ is the same as that of $\calLz$. 
Equation~\eqref{u-ph} is rewritten in the form 
\begin{equation}
u_\varphi=-\calQM c\rg\sin\vartheta\tan\vartheta. \label{u-QM}
\end{equation}
As expected,  the direction of rotational motion of the charged test particle is determined by the sign of $\calQM$ 
which is proportional to the product of $q$ and $\QM$. 

We have, from Eqs.~\eqref{K-def}, \eqref{dimless-L}, \eqref{ell-def} and \eqref{eq:C0}, 
\begin{equation}
\tan^2\vartheta=\frac{\calK^2}{\calQM^2}. \label{theta-ell-QM}
\end{equation}
By using Eq.~\eqref{QB-relation}, the dimensionless magnetic charge parameter is estimated to be
\begin{equation}
\calQM=\frac{q\QM}{4\pi \epzero mc^2\rg}=3.8\times 10^{6} \left(\frac{N^5}{N-1}\right)^{1\over2}\left(\frac{q}{e}\right)\left(\frac{\mpr}{m}\right)
\left(\frac{M}{4\times10^6 M_\odot}\right)
\left(\frac{B_N}{10^{-3}{\rm T}}\right)
\label{calQM-estimate}
\end{equation}
where $e$ and $\mpr$ are the elementary charge and the mass of a proton, respectively. 
In the neighborhood of Sgr~A*, $\calQM$ of both the proton and the electron is very large. 
Thus, we find, from Eq.~\eqref{calQM-estimate}, that, in the neighborhood of Sgr~A* 
and in the neighborhood of the center in M87 galaxy, 
$|\vartheta|\ll1$ holds for the charged test particle of $\calK^2$ which is much less than $\calQM^2$.

\section{Electric charge acquired by the black hole in monopole magnetic field}

In this section, we consider the electrification of a static black hole in the monopole magnetic field, $\calQM\neq0$. 
It is determined by the effective potential $V(R;\calE,\calK^2,\calQE)$ defined as Eq.~\eqref{V-def} 
whether the charged test particle falls into the black hole. Since the effective potential $V(R;\calE,\calK^2,\calQE)$ 
is completely the same as that in the case without the magnetic field studied in NMYI paper \cite{paper-I}, 
the results on the electrification of a static black hole obtained in NMYI paper is also obtained in the present case.

We assume that charged test particles compose plasma which is in local thermal equilibrium, and that 
the velocity of a charged test particle is given by the Maxwell distribution
\begin{align}
F(\bm{v})&=\left(\frac{m}{2\pi \kB T}\right)^{3\over2}\exp\left(-\frac{m} {2\kB T}\bm{v}^2\right), \label{Maxwell}
\end{align}
where $\kB$ and $T$ are the Boltzmann constant and the temperature, whereas
$\bm{v}$ is the spatial components of the four velocity $u^a$ 
with respect to the orthonormal basis associated to the static observer 
\begin{align}
e_{(t)\mu}&=\left(-c\sqrt{f(r)},0,0,0\right), \\
e_{(r)\mu}&=\left(0,\frac{1}{\sqrt{f(r)}},0,0\right), \\
e_{(\theta)\mu}&=\left(0,0,r,0\right), \\
e_{(\varphi)\mu}&=\left(0,0,0,r\sin\theta\right).
\end{align}
Hence, we have
\begin{align}
v_{(r)}&=e_{(r)\mu} u^\mu=\frac{1}{\sqrt{f(r)}}\frac{dr}{d\tau}, \label{vr-def}\\
v_{(\theta)}&=e_{(\theta)\mu} u^\mu=r \frac{d\theta}{d\tau}=\frac{u_\theta}{r}, \\
v_{(\varphi)}&=e_{(\varphi)\mu} u^\mu=r\sin\theta\frac{d\varphi}{d\tau}=\frac{u_\varphi}{r\sin\theta}. \label{vp-def}
\end{align}
From Eq.~\eqref{K-def}, the following equality holds,
\begin{equation}
v_{(\theta)}^2+v_{(\varphi)}^2=\frac{c^2\calK^2}{R^2}. \label{v-L-relation}
\end{equation}
Note that the assumption of the Maxwell distribution is justified only if $\kB T\ll mc^2$ holds. 

We are interested in the electrification of an electrically neutral black hole but in a monopole magnetic field.  
Thus, first we consider the case with $\calQE=0$ and $\calQM\neq0$. In this case, the effective potential 
for the charged test particle, which is defined as Eq.~\eqref{V-def} becomes  
\begin{equation}
V(R;\calK^2,\calQE=0)=\sqrt{\left(1-\frac{1}{R}\right)\left(1+\frac{\calK^2}{R^2}\right)}.
\end{equation}
The effective potential $V$ takes extrema at the roots of the equation 
\begin{equation}
\dfrac{d V(R;\calK^2,\calQE=0)}{d R}=0,
\end{equation} 
or equivalently
\begin{equation} 
R^2-2\calK^2R+3\calK^2=0. \label{extrema-eq}
\end{equation}
The roots of this equation are given as
\begin{equation}
R=R_\pm:=\calK^2\left(1\pm\sqrt{1-\frac{3}{\calK^2}}\right).  \label{extrema}
\end{equation}
We find from this equation that, in the case of $\calK^2<3$, there is no extremum of $V(R;\calK^2,\calQE=0)$, 
and hence the test particles with ingoing initial velocity necessarily 
enter the black hole. 

In the case of $\calK^2\geq3$, the sufficient condition for the test particle to fall into the black hole is  
$v_{(r)}<0$ and $\calE>V(R_-;\calK^2,\calQE=0)$. The normalization condition for the four-velocity, Eq.~\eqref{n-condition}, is rewritten in the form,
\begin{equation}
\frac{v_{(r)}^2}{c^2}=\frac{\calE^2}{1-1/R}-1-\frac{\calK^2}{R^2}. \label{n-condition-1}
\end{equation}
Hence the condition $\calE>V(R_-;\calK^2,\calQE=0)$ is equivalent to 
\begin{equation}
\frac{v_{(r)}^2}{c^2}>\frac{V^2(R_-;\calK^2,\calQE=0)}{1-1/R}-1-\frac{\calK^2}{R^2}, \label{falling-condition-0}
\end{equation}
where, from Eq.~\eqref{extrema}, we have
\begin{equation}
V^2(R_-;\calK^2,\calQE=0)=1+\frac{\calK^2}{27}\left[2-\frac{9}{\calK^2}+2\left(1-\frac{3}{\calK^2}\right)^{3\over2}\right].
\end{equation}
Using this expression, Eq.~\eqref{falling-condition-0} is rewritten as
\begin{equation}
2\calK^2\left(1-\frac{3}{\calK^2}\right)^{3\over2}<9+27\left(1-\frac{1}{R}\right)\frac{v_{(r)}^2}{c^2}
-\frac{27}{R}-\left[2-\frac{27}{R^2}\left(1-\frac{1}{R}\right)\right]\calK^2. \label{falling-condition-1}
\end{equation}
Since the L.H.S. of this inequality is non-negative, the R.H.S. should be non-negative so that there exists $\calK^2$ which satisfies this 
inequality. Since we focus on the case of $R\gg1$ initially, the following inequality should hold so that Eq.~\eqref{falling-condition-1} holds: 
\begin{equation}
\calK^2\leq \frac{9}{2}\left[1+3\left(1-\frac{1}{R}\right)\frac{v_{(r)}^2}{c^2}-\frac{3}{R}\right]\left[1-\frac{27}{2R^2}\left(1-\frac{1}{R}\right)\right]^{-1}. \label{falling-condition-2}
\end{equation}
Assuming Eq.~\eqref{falling-condition-2}, by taking the square of both sides of Eq.~\eqref{falling-condition-1}, 
a bit complicated manipulation leads 
\begin{equation}
c_0+c_1 \calK^2+c_2\calK^4+c_3\calK^6>0, \label{falling-condition-3}
\end{equation}
where
\begin{align}
c_0&=4, \\
c_1&=-1+9\left[\left(1-\frac{1}{R}\right)\frac{v_{(r)}^2}{c^2}-\frac{1}{R}\right]
\left\{2+3\left[\left(1-\frac{1}{R}\right)\frac{v_{(r)}^2}{c^2}-\frac{1}{R}\right]\right\}, \\
c_2&=-4\left[\left(1-\frac{1}{R}\right)\frac{v_{(r)}^2}{c^2}-\frac{1}{R}\right]
+\frac{18}{R^2}\left(1-\frac{1}{R}\right)\left\{1+3\left[\left(1-\frac{1}{R}\right)\frac{v_{(r)}^2}{c^2}-\frac{1}{R}\right]\right\}, \\
c_3&=-\frac{1}{R^2}\left(1-\frac{1}{R}\right)\left[4-\frac{27}{R^2}\left(1-\frac{1}{R}\right)\right].
\end{align}
By solving Eq.~\eqref{falling-condition-3} with respect to $\calK^2$, we obtain the following inequality: 
\begin{equation}
\calK^2<\calK_{\rm max}^2\left(v_{(r)}^2,R\right), \label{falling-condition-4}
\end{equation}
where, in the case that  $v_{(r)}^2/c^2$ and $1/R$ are the same order as one another and much less than unity,\footnote{We assumed 
$v_{(r)}^2/c^2 \gg 1/R$ in NMYI paper. The assumption in NMYI paper is equivalent to assume that the test particles are gravitationally unbound. 
By contrast, in this paper, we do not assume so, although the main result Eq.~\eqref{main-result} is the same as that in NMYI paper.} 
we have
\begin{equation}
\calK_{\rm max}^2\left(v_{(r)}^2,R\right)=4+8\left(\frac{v_{(r)}^2}{c^2}-\frac{1}{R}\right)
+{\cal O}\left( \frac{v_{(r)}^4}{c^4},\frac{1}{R^2}\right). \label{calKmax}
\end{equation}
Note that, in the case of $R\gg1$ and $v^2_{(r)}/c^2\ll1$, Eq.~\eqref{falling-condition-2} holds, if Eq.~\eqref{falling-condition-4} holds. 

By using Eq.~\eqref{v-L-relation}, we rewrite $v_{(\theta)}$ and $v_{(\varphi)}$ as
\begin{align}
v_{(\theta)}&=\frac{c\calK}{R}\cos\phi, \\
v_{(\varphi)}&=\frac{c\calK}{R}\sin\phi,
\end{align}
where $0\leq\phi<2\pi$. 

A test particle which initially moves outward will fall into the black hole only if it is bounded. 
As shown in Appendix \ref{Bound-particle}, the probability of a bounded particle initially moving outward for falling into the black hole 
is much less than that of a particle initially moving inward. Hence hereafter we consider only the probability $P$ of a particle 
initially moving inward for falling into the black hole.  It is given as
\begin{align}
P&=\int_{-\infty}^0 dv_{(r)}\left(\frac{m}{2\pi\kB T}\right)^{1\over2}\exp\left(-\frac{m}{2\kB T}v_{(r)}^2\right) \nonumber \\
&~~\times\int_0^{2\pi}d\phi\int_0^{\calK_{\rm max}^2\left(v_{(r)}^2,R\right)}d\calK^2\frac{mc^2}{4\pi\kB T R^2}\exp\left(-\frac{mc^2}{2\kB T R^2}\calK^2\right) 
\nonumber \\
&=\int_{-\infty}^0 dv_{(r)}\left(\frac{m}{2\pi\kB T}\right)^{1\over2}\exp\left(-\frac{m}{2\kB T}v_{(r)}^2\right) 
\left\{1-\exp\left[-\frac{mc^2}{2\kB T R^2}\calK_{\rm max}^2\left(v_{(r)}^2,R\right)\right]\right\} \nonumber \\
&\simeq\frac{mc^2}{\kB TR^2},  \label{main-result}
\end{align}
where we have used Eq.~\eqref{calKmax} in the last equality. 

Hereafter the test particle is assumed to be a proton or an electron. 
Any quantity of a proton is denoted by the character with a subscript ``p", whereas 
that of an electron is denoted by the character with a subscript ``e". 
Then the excess of the probability of the proton to enter the black hole is given as
\begin{equation}
\Delta P=P_{\rm p}-P_{\rm e}\simeq \frac{1}{R^2}\left(\frac{\mpr c^2}{\kB\Tp}-\frac{\mel c^2}{\kB\Te}\right).
\label{Number-n}
\end{equation}
If $\mpr/\Tp >\mel /\Te$ holds, $\Delta P$ is positive and the black hole will be positively electrified.  
By contrast, if $\mpr/\Tp < \mel/\Te$ holds, $\Delta P$ is negative and the black hole will be negatively electrified. 
If $\mpr/\Tp =\mel/\Te$ holds, no electrification of the black hole will occur. 
If the temperature of the protons and that of the electrons are identical to each other,  the black hole will be positively electrified 
because of $\mpr\gg\mel$. This result is completely the same as that obtained in NMYI paper. 

Although the detailed derivation is not shown in this paper, 
the maximal charge acquired by the black hole in the monopole magnetic field is also the same as 
that revealed in NMYI paper, i.e., $\calQE=1/2$ for a static black hole with no magnetic field. 
This value is not so large but the radius of ISCO of the charged test particle 
can be very different from that of the electrically neutral particle. 

It is easy to see that even in the case that the magnetic field is a split monopole, the present result is unchanged, 
since there is no magnetic charge $\calQM$ in the effective potential Eq.~\eqref{V-def}. 

\section{Collisionless plasma hovering above the black hole}\label{Heating}

For simplicity, in this section, we focus on the case of the electrically charge neutral black hole, $\calQE=0$, although 
the effect of the electric charge acquired by the black hole is an interesting issue.  
As shown in the subsection \ref{A-motion}, 
a charged test particle does not move on an equatorial plane 
but on a very thin cone. 

 \subsection{``Temperature" of bounded collisionless plasma}
 
\begin{figure}[!h]
\centering\includegraphics[width=9cm]{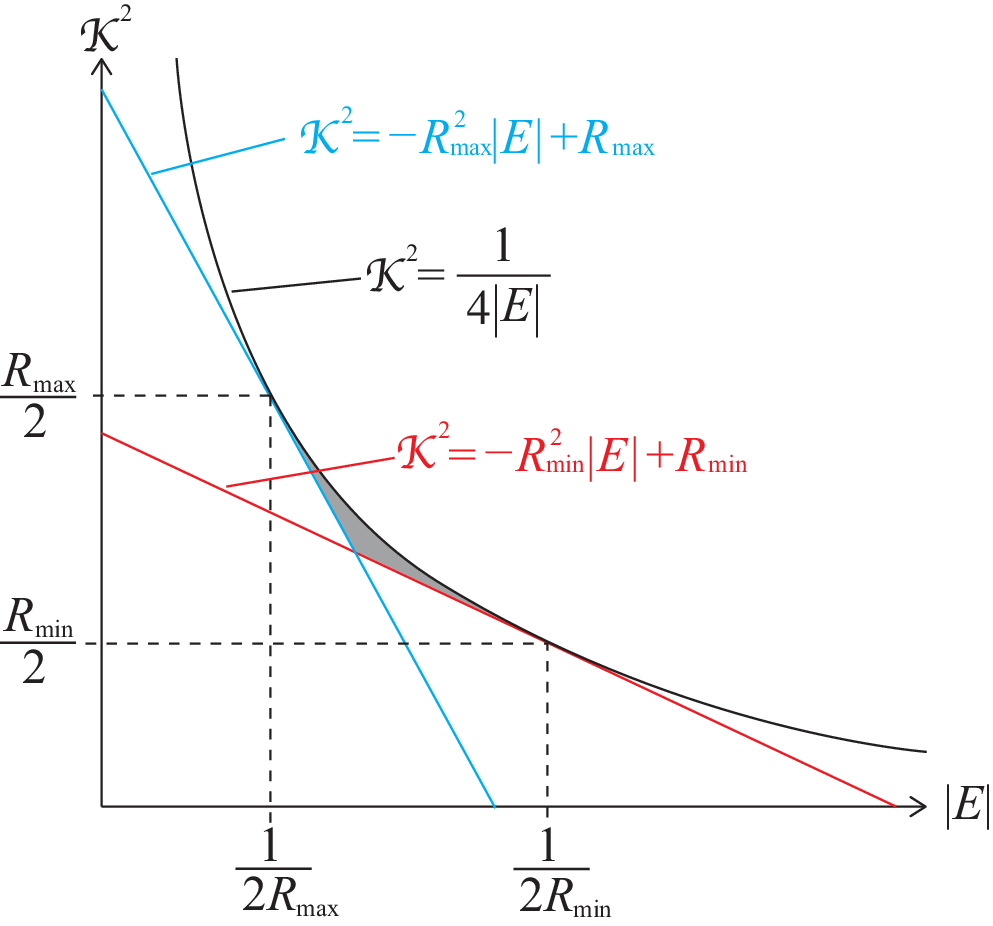}
\caption{ A schematic diagram. Equation \eqref{allowed-domain} is satisfied in the shaded region.  }
\label{fig:allowed-domain}
\end{figure}

Now, let us consider a lump of collisionless plasma bounded by 
gravity of the central black hole, which is assumed to be far from the black hole in the range, 
\begin{equation}
1\ll\Rmin<R<\Rmax. \label{B-domain}
\end{equation} 
The energy density of the plasma is assumed to be much less than that of the magnetic field and hence 
too small to affect the spacetime geometry. Thus the charged particles in the plasma can be treated as test particles. 
In such a situation, the energy equation for a test particle, which is derived from the normalization condition of the four-velocity,  
is approximately given as 
\begin{equation}
\left(\frac{dR}{d{\cal T}}\right)^2-\frac{1}{R}+\frac{\calK^2}{R^2}=E, \label{Newton-E-eq}
\end{equation}
where $E={\cal E}^2-1$. Iff   
\begin{equation}
-\frac{1}{4\calK^2}\leq E < 0 \label{Reality-0}
\end{equation}
holds, the motion of the test particle is bounded in the domain
\begin{equation}
R_-\leq R\leq R_+,
\end{equation}
where
\begin{equation}
R_\pm=-\frac{1}{2E}\left(1\pm\sqrt{1+4E\calK^2}\right).
\end{equation}
Then, from Eq.~\eqref{B-domain}, the allowed $E$ and $\calK^2$ are determined by
\begin{equation}
\Rmin\leq R_- \leq R_+\leq\Rmax. \label{allowed-domain}
\end{equation}
From Eq.~\eqref{Reality-0}, the inequality, $E<0$, should hold. 
The condition $\Rmin\leq R_-$ leads to two inequalities,
\begin{equation}
|E|\leq \frac{1}{2\Rmin} \label{allowed-1}
\end{equation}
and
\begin{equation} 
\calK^2\geq -\Rmin^2|E|+\Rmin,
\end{equation}
whereas the condition $R_+\leq\Rmax$ leads to
\begin{equation}
|E|\geq\frac{1}{2\Rmax} \label{allowed-2}
\end{equation}
and
\begin{equation}
\calK^2\geq -\Rmax^2|E|+\Rmax.
\end{equation}
An inequality
\begin{equation}
\calK^2\leq\frac{1}{4|E|} \label{Reality-1}
\end{equation}
is derived from Eq.~\eqref{Reality-0}. 
Inequalities \eqref{allowed-1}--\eqref{Reality-1} determine the allowed $|E|$ and $\calK^2$ 
(see a schematic diagram in Fig.~\ref{fig:allowed-domain}). 

In the situation we consider, the square of the dimensionless velocity, $\calV^2$, is given by
\begin{align}
\calV^2=\left(\frac{dR}{d{\cal T}}\right)^2+\frac{\calK^2}{R^2}. \label{calV-def}
\end{align}
Since the motion of the test particle is bounded, the time average of $(dR/d{\cal T})^2$ is given as
\begin{align}
\left\langle \left(\frac{dR}{d\calT}\right)^2\right\rangle&=\lim_{T\rightarrow \infty}
 \frac{1}{2T}\int_{-T}^{T}\left(\frac{dR}{d\calT}\right)^2 d\calT \nonumber \\
&=\lim_{T\rightarrow\infty} \frac{1}{2T}
\int_{-T}^{T}\left[\frac{d}{d\calT}\left(R\frac{dR}{d\calT}\right)-R\frac{d^2R}{d\calT^2}\right] d\calT \nonumber \\
&=\lim_{T\rightarrow \infty} \frac{1}{2T}
\int_{-T}^{T}\left(\frac{1}{2R}-\frac{\calK^2}{R^2}\right) d\calT \nonumber \\
&=\left\langle\frac{1}{2R}\right\rangle-\left\langle\frac{\calK^2}{R^2}\right\rangle, 
\label{average}
\end{align}
where, in the third equality, we have used the equation of motion derived by differentiating Eq.~\eqref{Newton-E-eq} with respect to $\calT$,
\begin{equation}
\frac{d^2R}{d\calT^2}=-\frac{1}{2R^2}+\frac{\calK^2}{R^3}.
\end{equation}
From Eqs.~\eqref{calV-def} and \eqref{average}, we have 
\begin{equation}
\langle\calV^2\rangle=\left\langle\frac{1}{2R}\right\rangle,
\end{equation}
or equivalently, 
\begin{equation}
\langle\calV^2\rangle=-E. 
\end{equation}
From the condition Eq.~\eqref{allowed-domain}, \eqref{allowed-1} and \eqref{allowed-2}, we obtain
\begin{equation}
\frac{1}{2\Rmax}\leq\langle \calV^2\rangle \leq \frac{1}{2\Rmin}.
\end{equation}
This inequality implies that if the plasma is confined in the domain $\Rmin\leq R\leq\Rmax$, 
the time average of its kinetic energy is restricted by
\begin{equation}
\frac{mc^2}{4\Rmax}\leq \left\langle \frac{1}{2}mc^2\calV^2 \right\rangle \leq \frac{mc^2}{4\Rmin}.
\end{equation}  
In the case of the proton, the time average of the kinetic energy is bounded as 
\begin{equation}
2.35\times10^{6}\left(\frac{100}{\Rmax}\right){\rm eV}\leq \left\langle\frac{1}{2}\mpr c^2 \calV^2\right\rangle \leq 2.35\times10^{7}\left(\frac{10}{\Rmin}\right){\rm eV}.\label{k-energy}
\end{equation}
The velocities of protons and electrons on the same orbit are equal to each other since they are governed by gravity of the black hole. 
Thus the average of the kinetic energy of protons is about 1833 times larger than 
that of an electron. 

Let us consider a lump of collisionless plasma composed of protons and electrons, whose size is assumed to be, for example, the order of the gravitational radius $\rg$. Its radial extent, $\Rmax-\Rmin$, is of order one and the tangential extent is $2\Rmin\Theta=1$. Here note that the half apex angle $\vartheta$ of the cone on which the motion of the charged test particle is restricted is much less than $\Theta$ in the situation of our interest, $10\lesssim\Rmin<\Rmax\lesssim 100$ (see Eqs.~\eqref{theta-ell-QM} and \eqref{calQM-estimate}). This means that the Larmor radius  of the charged test particle is much less than the size of the plasma lump, since the Larmor radius is less than $\Rmax\sin\vartheta$. Hence the random motions of the charged test particles in the plasma lump are Larmor one if the plasma is really collisionless. In this case, the time average of the kinetic energy estimated in Eq.~\eqref{k-energy} might be regarded as the ``temperature" of the plasma lump. 
Here it is very remarkable fact that the kinetic energy of a charged test particle  
is equal to that of the ``Keplerian" motion, 
since the energy equation of the charged test particle is given by Eq.~\eqref{Newton-E-eq}.  Hence, the temperature of the plasma lump can be very high in the neighborhood of the black hole in the monopole magnetic field. The origin of this temperature is gravity. 

As mentioned, in the case that the plasma is composed of protons and electrons, their kinetic energies should be very different from each other due to the mass difference. Even if the velocity distribution of the protons and that of the electrons become Maxwell ones with the same temperature, 
the domain allowed for the motions of protons and those of electrons will be very different. Hence, the local thermal equilibrium seems to be unachievable in the plasma lump hovering on the black hole. 

It should be noted that the average velocity of electrons will be equal to that of protons in a plasma lump. This situation is very different from the situation assumed in the discussion of the electrification of a static black hole of the NHYI paper \cite{paper-I}, where we have assumed that the velocity distribution of electrons and that of protons are Maxwell one with not so different temperatures from each other. Hence, we should note that lumps of collisionless plasma will not be suppliers of electric charge to the black hole.

Here note that the global monopole magnetic field is not a necessary condition for the hovering motion. If the configuration of the magnetic field only in the domain of the motion of the charged test particle is locally similar to the monopole magnetic field, the charged test particle will 
hover on the black hole. In the stationary axisymmetric magnetosphere, the neighborhood of the rotational symmetry axis of a rotating black hole may be the case (see for example, Fig.~3 in Ref.~\cite{MG2004}). Thus, in the realistic situation like the neighborhood of Sgr~A*,  
the heating of a plasma lump may occur in the vicinity of the symmetry axis of rotating black hole.  

 \subsection{Collisionless condition}

The two-body relaxation time for the plasma composed of protons and electrons 
is estimated as the time scale for which the accumulated changes in velocity 
through  multiple two-body Coulomb scatterings becomes the same order of 
the initial velocity. The shortest time scale is given by the electron-electron scattering,  
\begin{equation}
T_{\rm relax}=\frac{4\pi\epzero^2 \mel^2 \ell^3 v_{\rm e}^3}{e^4\ln(\lambda_{\rm D}/\ell)} ,
\end{equation}
where $\ell$, $v_{\rm e}$ and $\lambda_{\rm D}$  are the mean separation of nearest two electrons, i.e., 
$(n/2)^{-1/3}$, the mean velocity of electrons and the Debye length (see, for example, Appendix B in NMYI paper). 
By contrast, the period of the Larmor motion in the present situation is given by
\begin{equation}
T_{\rm Larmor}=\frac{2\pi r\sin\vartheta}{v_{(\varphi)}}\simeq \frac{2\pi \sqrt{3}r\sin\vartheta}{v_{\rm e}}.
\end{equation}
The ratio of these two time scales is given by
\begin{equation}
\frac{T_{\rm relax}}{T_{\rm Larmor}}
=\frac{2\epzero^2 \mel^2 v_{\rm e}^4}{\sqrt{3}\hskip0.05cm e^4 n r\sin\vartheta\ln(\lambda_{\rm D}/\ell)}.
\end{equation}
By assuming the isotropy of the velocity distribution, we have, form Eq.~\eqref{K-def}, 
\begin{equation}
v_{\rm e}^2=v_{(r)}^2+v_{(\theta)}^2+v_{(\varphi)}^2\simeq 3v_{(\varphi)}^2=\frac{3c^2\calK^2}{2R^2}.
\end{equation}
From Eq.~\eqref{theta-ell-QM}, we have
\begin{equation}
\sin\vartheta=\frac{\calK}{\sqrt{\calQM^2+\calK^2}}. 
\end{equation}
We find from Fig.~\ref{fig:allowed-domain} that $R_{\rm min}/2<\calK^2<R_{\rm max}/2$ holds if we focus on 
particles bounded within $R_{\rm min}<R<R_{\rm max}$. We 
consider the case of $10<R_{\rm min}<R_{\rm max}<100$, 
and hence we should assume $5<\calK^2<50$. 
Then, from Eq.~\eqref{calQM-estimate}, we have $\calQM^2\gg\calK^2$, and hence we may approximate $\sin\vartheta$ as
\begin{equation}
\sin\vartheta\simeq \frac{\calK}{|\calQM|}.
\end{equation}
Thus, we have
\begin{align}
\frac{T_{\rm relax}}{T_{\rm Larmor}}
&=\frac{9\epzero^2 \left(\mel c^2\right)^2}{2\sqrt{3}\hskip0.05cm e^4}\times \frac{\calK^3|\calQM|}{\rg n R^5 \ln(\lambda_{\rm D}/\ell)}
\nonumber \\
&=\frac{9\epzero^2 c^3 \mel \calK^3}{2\sqrt{3}e^3 nR^5\ln(\lambda_{\rm D}/\ell)}\left(\frac{N^5}{N-1}\right)^{1\over2} 
\left(\frac{|q|}{e}\right)
\left(\frac{\mel}{m}\right)|B_N|,
\end{align}
where we have used Eq.~\eqref{calQM} with Eq.~\eqref{QB-relation} in the second equality.

If $T_{\rm relax} \gg T_{\rm Larmor}$, or equivalently,
\begin{align}
n&\ll \frac{9\epzero^2 c^3 \mel \calK^3}{2\sqrt{3}e^3 R^5\ln(\lambda_{\rm D}/\ell)}\left(\frac{N^5}{N-1}\right)^{1\over2} 
\left(\frac{|q|}{e}\right)
\left(\frac{\mel}{m}\right)|B_N|
\nonumber \\
&\simeq\frac{1.52\times 10^{22}}{\ln(\lambda_{\rm D}/\ell)}
\left(\frac{100\rg}{r}\right)^5\left(\frac{N}{10}\right)^2\left(\frac{\calK}{5}\right)^3\left(\frac{|q|}{e}\right)
\left(\frac{\mel}{m}\right)\left(\frac{|B_N|}{10^{-3}{\rm T}}\right)[{\rm m}^{-3}]
\end{align}
holds, the plasma can be regarded to be collisionless, where, in the last equality, we have assumed $N\gg1$.  
This condition is satisfied in the situations of our interest.

 \subsection{Magnetic field produced by the electric current caused by charged test particles}\label{Current-1}

\begin{figure}[!h]
\centering\includegraphics[width=9cm]{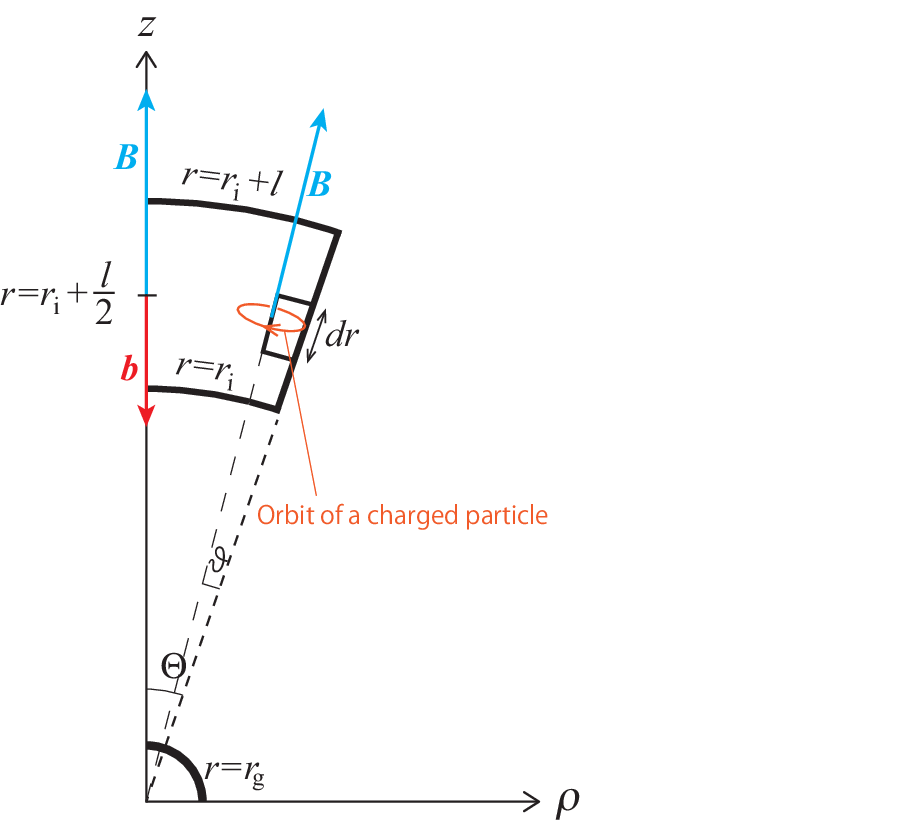}
\caption{ A clump of plasma hovering on a black hole is depicted, where $\rho:=\sqrt{x^2+y^2}$. 
It occupies a domain $r_{\rm i}\leq r\leq r_{\rm i}+l$ and $0\leq\theta\leq\Theta$.}

\label{fig:hovering-lump}
\end{figure}

The electric current is caused by the motions of charged particles so as 
to cancel out the background monopole magnetic field. If the electric current is too large, 
the hovering motions in the plasma lump is highly affected. 
Thus we consider a lump of charged particles hovering on a black hole and estimate the magnetic field generated by 
the motion of the charged particles. In order to simplify the calculations, 
we assume that the lump occupies a domain specified by 
$r_{\rm i}\leq r\leq r_{\rm i}+l$ and $0\leq\theta\leq\Theta$ with 
\begin{equation}
2r_{\rm i}\sin\Theta=l
\end{equation} 
(see Fig.~\ref{fig:hovering-lump}). 
A particle with $\calK$ moves on a cone with a half apex angle $\vartheta$ given by Eq.~\eqref{theta-ell-QM}. 
The electric currents generated by charged test particles cancel out inside the lump, and hence the electric current on the surface 
$\theta=\Theta$ remains. 

We assume that the  system is static and axisymmetric, or equivalently, the four-vector potential $A_a$ does not 
depend on $t$ and $\varphi$, and $j^\alpha\propto \delta_3^\alpha$.  
Then, Maxwell's equations are given as 
\begin{align}
\sum_{i,j=1}^3\frac{\partial}{\partial x^i}\left(\sqrt{-g}g^{00}g^{ij}\frac{\partial A_0}{\partial x^j}\right)&=0, \\
\frac{\partial }{\partial \theta}\left(\sqrt{-g} F^{21}\right)&=0, \\
\frac{\partial }{\partial r}\left(\sqrt{-g} F^{12}\right)&=0, \\
\frac{1}{r^2}\frac{\partial }{\partial r}\left(r^2 F^{13}\right)+\frac{1}{\sin\theta}\frac{\partial}{\partial\theta}\left(\sin\theta F^{23}\right)&=-\frac{4\pi}{c^2\epzero}j^3.
\end{align}
If $A_0$, $A_1$ and $A_2$ are constant, the first, second and third equations are trivially satisfied, and we adopt these trivial solutions. 
The forth equation is rewritten in the form
\begin{equation}
\frac{\partial}{\partial r}\left(f(r)\frac{\partial A_\varphi}{\partial r}\right)
+\frac{\sin\theta}{r^2}\frac{\partial }{\partial\theta}\left(\frac{1}{\sin\theta}\frac{\partial A_\varphi}{\partial\theta}\right)
=-\frac{4\pi}{c^2\epzero}j_\varphi. \label{Ampere-law}
\end{equation}
We assume the domain of $r\gg\rg$, or equivalently, $f(r)\simeq1$.  Then, Eq.~\eqref{Ampere-law} becomes Ampere's law 
in the flat spacetime. Since our purpose is the 
order estimate of the magnetic field generated by the motion of charged particles, 
this assumption will not lead to serious incorrectness.  
 
Then a circular electric current $dI$ going through a domain $r\in [r,r+dr]$, $\theta\in[\Theta-\vartheta,\Theta]$ 
and $\varphi\in[0,2\pi]$ is given as 
\begin{equation}
dI=\frac{1}{2}q n v_{(\varphi)} r\vartheta dr,
\end{equation}
where $n$ is the number density of the charged particles with electric charge $q$, and $v_{(\varphi)}$ is defined as 
Eq.~\eqref{vp-def} (also see Fig.~\ref{fig:hovering-lump} and Appendix \ref{Current-2}). 

Then the magnetic field generated at the center of this clump, 
$(r,\theta)=\left(r_{\rm i}+l/2,0\right)$, by the circular electric current is parallel to the $z$-axis and its $z$-component $db^z$ is given as
\begin{equation}
db^z=\frac{dI r^2\sin^2\Theta}{2c^2\epzero}\left[r^2\sin^2\Theta+\left(r_{\rm i}+\frac{l}{2}-r\cos\Theta\right)^2\right]^{-3/2}
\end{equation}
We need to know the  upper bound of the field strength generated by the motion of charged particles in the lump. 
Since the maximal absolute value of $v_{(\varphi)}$ with the energy fixed is realized in the case that the charged particles moves circularly, 
we assume the circular motion of the charged particles. In the case of the circular motion, the orbital radius is the minimum of the effective potential $V(R;\calK,\calQE=0)$ 
defined as Eq.~\eqref{V-def}, and hence, from Eq.~\eqref{extrema-eq}, we have
\begin{equation}
\calK^2=\frac{R^2}{2R-3} \label{circular-calK}
\end{equation}
From this equation and Eq.~\eqref{theta-ell-QM}, we have
\begin{equation}
\vartheta=\arctan \frac{\calK}{|\calQM|}=\arctan \left(\frac{R}{|\calQM|\sqrt{2R-3}}\right).
\label{circular-theta-ell-QM}
\end{equation}
From Eqs.~\eqref{u-QM}, \eqref{vp-def} and \eqref{circular-theta-ell-QM}, we have
\begin{equation}
v_{(\varphi)}=-\frac{c\calQM}{R}\tan\vartheta=-\frac{c~{\rm sgn}(\calQM)}{\sqrt{2R-3}}, \label{v-phi}
\end{equation}
where sgn$(x)$ is the sign function. By assuming $0<\vartheta\ll1$ or equivalently, $|\calQM|\gg R$, since  
\begin{align}
\vartheta &\simeq \frac{R}{|\calQM|\sqrt{2R-3}}
\end{align}
is a good approximation, we have
\begin{equation}
b^z=-\frac{nq\rg\sin^2\Theta}{4c\epzero\calQM}
\int_{R_{\rm i}}^{R_{\rm i}+L}\frac{R^4dR}{(2R-3)}\left[R^2-2R\left(R_{\rm i}+\dfrac{L}{2}\right)\cos\Theta+\left(R_{\rm i}+\dfrac{L}{2}\right)^2\right]^{-3/2}, \label{b-field-0}
\end{equation}
where $R_{\rm i}=r_{\rm i}/\rg$ and $L=l/\rg$. 
By replacing the integration variable $R$ by $\beta$ defined as
\begin{equation}
R=R_{\rm i}\left(1+\frac{L}{2R_{\rm i}}\right)\left[\frac{L}{2R_{\rm i}}\sinh\beta+\sqrt{1-\left(\frac{L}{2R_{\rm i}}\right)^2}\right],
\end{equation}
and by assuming $L\ll R_{\rm i}$, we obtain
\begin{equation}
 b^z=-\frac{nq\rg}{4c\epzero\calQM}\times\frac{R_{\rm i}^2}{\left(2R_{\rm i}-3\right)}\int^{\ln(\sqrt{2}+1)}_{\ln(\sqrt{2}-1)}\frac{d\beta}{\cosh^2\beta}\left[1+{\cal O}\left(\frac{L}{R_{\rm i}}\right)\right]
 \simeq -\frac{\sqrt{2}\hskip0.05cm nq\rg R_{\rm i}}{8\hskip0.05cm c\epzero\calQM},
\end{equation}
where in the last equality, we have used the assumption of $R_{\rm i}\gg1$. 
By assuming $R_{\rm i}\gg1$ and $N\gg1$, we have
\begin{align}
\left|\frac{b^z}{B^r|_{r=r_i}}\right|&\simeq\frac{\sqrt{2}\hskip0.05cm nq\rg R_{\rm i}}{8\hskip0.05cm c\epzero |\calQM|}\times\frac{R_{\rm i}^2}{N^2 B_N}
\simeq \frac{\sqrt{2}\hskip0.05cm nmR_{\rm i}^3}{8\epzero B_N^2 N^4},
\end{align}
where we have used Eqs.~\eqref{QB-relation} and \eqref{calQM} in the last equality. 
If $|b^z/B^r|_{r=r_i}|\ll1$, or equivalently, 
\begin{align}
 n&\ll \frac{4\sqrt{2}\hskip0.05cm\epzero B_N^2 N^4}{mR_{\rm i}^3}
 \nonumber\\
&\simeq 2.99\times10^8 \left(\frac{\mpr}{m}\right)\left(\frac{100\rg}{r_{\rm i}}\right)^3
\left(\frac{B_N}{10^{-3}{\rm T}}\right)^2 \left(\frac{N}{10}\right)^4
 [{\rm m}^{-3}]
\end{align}
holds, our analysis can be valid. This condition is not satisfied in broad-line regions in typical AGN\cite{Peterson} 
but a typical HII region.

 \subsection{Backreaction due to the electromagnetic radiation}\label{sec:Radiation-Reaction}

From the results obtained in Sec.~\ref{A-motion}, we can estimate the frequency of the electromagnetic radiation generated by 
the charged test particle. The frequency $\nu$ of the angular motion of the test particle is estimated to be 
\begin{align}
\nu&=\frac{1}{2\pi}\left|\frac{d\varphi}{d\tau}\right| = \frac{\left|u_\varphi\right|}{2\pi r^2\sin^2\vartheta} 
\nonumber \\
&=\frac{c}{2\pi \rg}\sqrt{\calQM^2+\calK^2}\left(\frac{\rg}{r}\right)^2 \nonumber \\
&\simeq2.8\times10^5\left(\frac{100\rg}{r}\right)^2\left(\frac{N}{10}\right)^2\left(\frac{q}{e}\right)\left(\frac{\mel}{m}\right)
\left(\frac{B_N}{10^{-3}{\rm T}}\right){\rm Hz},
\label{f-estimate}
\end{align}
where we have used Eqs.~\eqref{eq:C0}, \eqref{u-Lz} and \eqref{theta-ell-QM} in the third equality  
and Eq.~\eqref{calQM-estimate} with the replacement of $\mpr$ by $\mel$ in the last equality, and 
$\calQM^2\gg\calK^2$ and $N\gg1$  have also been assumed in the last equality. 
From this result, we find that the radiation of very low frequencies may be generated 
by electrons in this system, if the strength of the magnetic field is the same order as that in the neighborhood of Sgr~A* 
or of the center of M87 galaxy.   

The rotational velocity of the circular motion is given by
\begin{equation}
v_{(\varphi)}^2=\frac{c^2\calK^2}{R^2}=\frac{c^2}{2R-3},
\end{equation}
where we have used Eq.~\eqref{circular-calK} in the last equality.
Hence, the rotational velocity is much less than the speed of light if the radial position of the test particle satisfies $r\gg\rg$.  

In the case that the motion of the charged test particle is non-relativistic, 
the radiation power $I_{\rm EM}$ of the magnetic bremsstrahlung by one particle is estimated by the dipole formula
\begin{equation}
I_{\rm EM}=\frac{1}{6\pi c^3\epzero}\sum_{i=1}^3\frac{d^2D^i}{dt^2}\frac{d^2D^i}{dt^2}, \label{q-formula}
\end{equation}
where $D^i$ is the electric dipole moment defined as
\begin{equation}
D^i=qx^i \label{d-def}
\end{equation}
with the spatial trajectory of the charged test particle, $x^i=x^i(t)$,  which is approximately given as 
\begin{align}
x^1&=r\sin\vartheta\cos\left(2\pi \nu t\right), \\
x^2&=r\sin\vartheta\sin\left(2\pi \nu t\right), \\
x^3&=r\cos\vartheta.
\end{align}

By assuming that the radial velocity is much smaller compared to the angular velocity, $\left|\dfrac{dr}{dt}\right|\ll \nu r$, we have  
\begin{equation}
I_{\rm EM}=\frac{(2\pi \nu)^4 q^2 r^2 \sin^2\vartheta}{6\pi c^3 \epzero}.
\end{equation} 
Then, by using Eq.~\eqref{f-estimate}, we have 
\begin{align}
I_{\rm EM}
&=\frac{ce^2}{6\pi\epzero \rg^2}\frac{\calK^2(\calQM^2+\calK^2)}{R^6}\left(\frac{q}{e}\right)^2
\nonumber \\
&\simeq \frac{e^4 B_N^2}{12\pi\epzero c\mel^2}\frac{N^4}{R^5}\left(\frac{\mel}{m}\right)^2\left(\frac{q}{e}\right)^4
\nonumber \\
&\simeq4.95\times10^{-8}\left(\frac{N}{10}\right)^4\left(\frac{100\rg}{r}\right)^5\left(\frac{\mel}{m}\right)^2\left(\frac{q}{e}\right)^4
\left(\frac{B_N}{10^{-3}{\rm T}}\right)^2 {\rm eV/s},  \label{D-radiation}
\end{align} 
where we have assumed $R\gg1$ and $\calQM^2 \gg \calK^2\simeq R/2$ [see Eq.~\eqref{circular-calK}].
If the released energy is equal to about 10$\%$ of the rest mass energy of the particle, the particle will fall into the black hole. 
The time scale for which the released energy even at $r=100\rg$ is about 10$\%$ of the rest mass energy of an electron is 
$1.03\times10^{13}$sec which is much shorter than the age of the Universe,  $1.38\times10^{10}$yr$=4.3\times10^{17}$sec.  
In the case of $r\simeq 10\rg$, the time scale is $10^5$ times shorter than the case of $r=100\rg$, i.e., $1.03\times 10^8$ sec $\simeq3.3$yr, 
and hence effects of the radiation on the motion of electrons around $r=10\rg$ can cause observable phenomena. 
By contrast, since $I_{\rm EM}$ of a proton is $(\mel/\mpr)^2=2.96\times10^{-7}$ times that of an electron, 
the time scale of a proton at $r=100\rg$ is $3.37\times10^6$ times longer than that of the electron, which is much longer than the age of the Universe. 
Even in the case of $r=10\rg$, the time scale is $1.1\times10^7$yr of which effect cannot be observed.  
The difference between the time scales of an electron and a proton may cause selective accretion 
of negative electric charges to the black hole from a collisionless plasma lump. 

\section{Summary and discussions}

We studied the motion of a charged test particle around a spherically symmetric black hole in a monopole magnetic field. 
Its radial motion is completely the same as that of the charged test particle in the case with no magnetic field. This fact immediately leads to 
the same results on the electrification of a black hole as the case with no magnetic field. If plasma composed of non-relativistic protons and 
electrons whose temperatures are not so different from each other surrounds a spherically symmetric 
black hole in the monopole magnetic field, selective accretions of protons will occur 
due to the difference of the masses between a proton and an electron. 
Since $\Te<\Tp<5\Te$ will hold in the situation of our interest\cite{Mos,Dexter,Yuan-Narayan}, 
the black hole will acquire positive charge. The total amount of charge will be a very small amount but its effect on the charged particles 
is very large. For example, the radius of ISCO of an electron can be one order larger than those of the electrically 
neutral test particle\cite{paper-I}. 
We need to take into account these effects on the plasma surrounding an electrified black hole. 

By contrast to the radial motion, the motion in angular directions of the charged test particles in the monopole 
magnetic field is very different from that in the case with no magnetic field. 
The motion of an electrically neutral test particle is restricted to an equatorial plane, 
whereas that of a charged test particle is restricted on a cone whose half apex angle $\vartheta$ 
is determined by the magnetic charge of the black hole, the electric charge and the angular momentum of the particle. 
In the situation with the magnetic field whose strength is similar to Sgr~A* or the center of M87 galaxiy, 
the apex angle $\vartheta$ of the cone is much less than one radian, if the specific angular momentum of the charged particle 
is much less than $10^6 c\rg$. Here note that the orbital radial position $r$ of a circularly moving particle 
with the specific angular momentum $K$ is given as $r\simeq K^2/c\rg$.   
Hence the particle bounded within the domain of $r<10^4\rg$ will have a specific angular momentum less than $10^2c\rg$.    
These results imply that a charged test particle of a circular motion at the radial position $r$ 
does not stay on an equatorial plane but hovers on a black hole in a very narrow region of the tangential extent, 
or in other words, the Larmor radius, $r \vartheta$, 
which is much less than $r$.  The distribution of plasma around the black hole in the monopole 
magnetic field may be rather different from the case with no magnetic field. 

By the same reason, a plasma lump can hover on the black hole in the monopole magnetic field, 
if the size of the lump is much larger than the 
Larmor radii of the constituent charged particles, $r\vartheta$.  
Here note that a plasma particle of the Larmor radius $r\vartheta$ has kinetic energy equal 
to that of the Keplerian motion of the orbital radius $r$. 
Due to this fact, the mean kinetic energy of plasma particles, or equivalently, 
the local temperature of the plasma lump may be extremely high.
For example, the temperature of protons may be the order of $10^{10}$K 
when the radial position of the plasma lump is equal to about $100\rg$ 
in the case of the black hole with the mass same as that of Sgr~A*.  
Although the temperature of the plasma lump is very high, the thermal radiation will not be emitted since the plasma 
is too diffuse to be in thermal equilibrium with photons in the situation we consider. 
It seems to be difficult to observe this high temperature of the plasma, even if it is the case. 

It is a crucial prediction that the average velocity of electrons will be equal to that of protons, or in other words, 
the temperature of electrons is much lower than that of protons, in a plasma lump. 
This situation is very different from the situation assumed in the discussion of the electrification of a static black hole \cite{paper-I}, 
where we have assumed that the velocity distribution of electrons and that of protons 
are Maxwell one with not so different temperatures from each other. Hence, we should note that 
bounded lumps of collisionless plasma will not be suppliers of positive electric charge to the black hole. 
However, as shown in Sec.~\ref{sec:Radiation-Reaction}, electrons in a plasma lump can be selectively accreted on the black hole 
through the energy loss due to the electromagnetic radiation. Thus, lumps of collisionless plasma bounded around a black hole 
may be suppliers of negative electric charges to the black hole in a monopole magnetic field.

\section*{Acknowledgments}
We are grateful to colleagues in the astrophysics and gravity group in Osaka Metropolitan University. 
This work was supported by JSPS KAKENHI Grants No. JP21K03557 (K.N.), No. JP21H05189 (H.Y.), No. JP22H01220 (H.Y.), 
and MEXT Promotion of Distinctive Joint Research Center Program JPMXP0723833165 (K.N. and H.Y.).

\appendix

\section{About a test particle initially moving outward}\label{Bound-particle}

\begin{figure}[!h]
\centering\includegraphics[width=9cm]{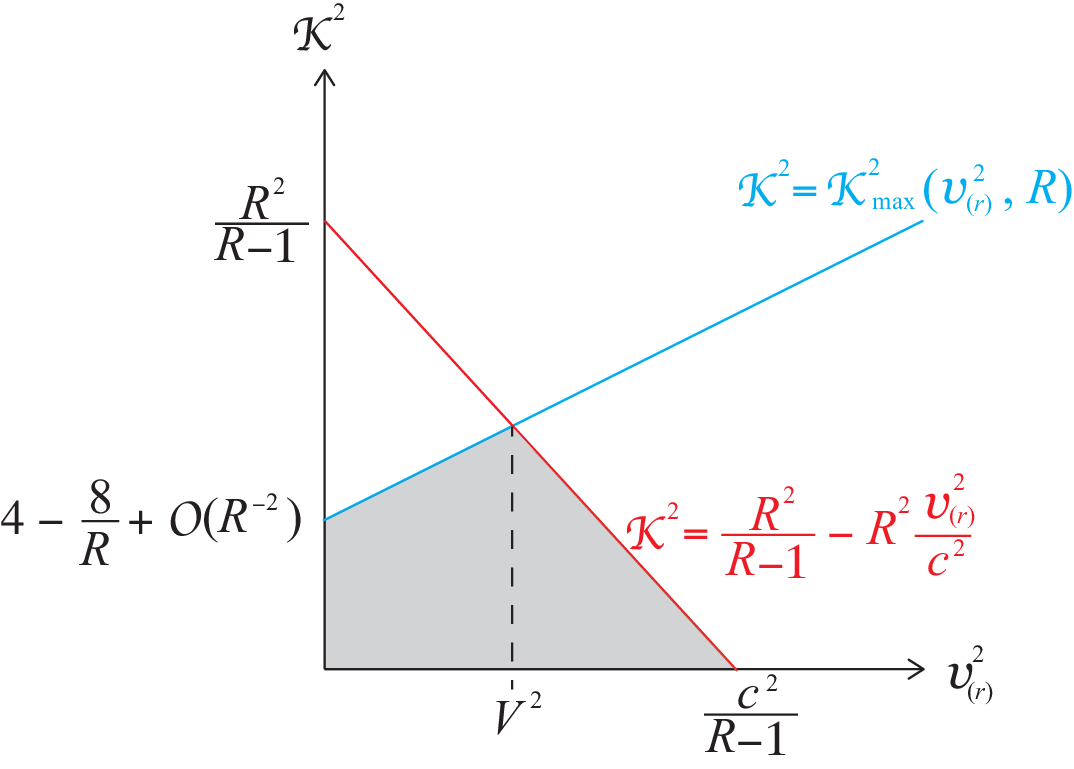}
\caption{ The parameter region in which both Eq.~\eqref{falling-condition-4} and Eq.~\eqref{falling-condition-5} are satisfied 
shown as a shaded domain in the $(v_{(r)}^2$,$\calK^2)$-plane. 
}
\label{fig:Integral-domain}
\end{figure}

We consider a test particle initially moving outward. It should be bounded, or equivalently, ${\cal E}^2<1$ holds so that it falls into the black hole.  
Equation~\eqref{n-condition-1} implies that the bounded test particle should satisfy
\begin{equation}
\frac{v_{(r)}^2}{c^2}<\frac{1}{1-1/R}-1-\frac{\calK^2}{R^2}, 
\end{equation}
or equivalently
\begin{equation}
\calK^2<R^2\left(\frac{1}{R-1}-\frac{v_{(r)}^2}{c^2}\right). \label{falling-condition-5}
\end{equation}
In Fig.~\ref{fig:Integral-domain}, we depict the parameter region in which both Eq.~\eqref{falling-condition-4} 
and Eq.~\eqref{falling-condition-5} are satisfied as a shaded domain in the $(v_{(r)}^2$,$\calK^2)$-plane, 
where $v_{(r)}^2=V^2$ is a root of 
\begin{equation}
\calK_{\rm max}^2(v_{(r)}^2,R)=R^2\left(\frac{1}{R-1}-\frac{v_{(r)}^2}{c^2}\right).
\end{equation} 
For $R\gg 1$, by using Eq.~\eqref{calKmax}, we have
\begin{equation}
V^2=\frac{c^2}{R}\left[1-\frac{4}{R}+{\cal O}\left(R^{-2}\right)\right].
\end{equation}

Thus, in the case of $R\gg1$, the probability $P_{\rm out}$ of the test particle initially moving outward falling into the black hole is given as
\begin{align}
P_{\rm out}&=\int_0^V dv_{(r)}\left(\frac{m}{2\pi\kB T}\right)^{1\over2}\exp\left(-\frac{m}{2\kB T}v_{(r)}^2\right) \nonumber \\
&~~\times\int_0^{2\pi}d\phi\int_0^{\calK_{\rm max}^2\left(v_{(r)}^2,R\right)}d\calK^2\frac{mc^2}{4\pi\kB T R^2}\exp\left(-\frac{mc^2}{2\kB T R^2}\calK^2\right) 
\nonumber \\
&+\int_V^{\frac{c}{\sqrt{R-1}}} dv_{(r)}\left(\frac{m}{2\pi\kB T}\right)^{1\over2}\exp\left(-\frac{m}{2\kB T}v_{(r)}^2\right) \nonumber \\
&~~\times\int_0^{2\pi}d\phi\int_0^{\frac{R^2}{R-1}-\frac{R^2v_{(r)}^2}{c^2}}d\calK^2\frac{mc^2}{4\pi\kB T R^2}\exp\left(-\frac{mc^2}{2\kB T R^2}\calK^2\right) 
\nonumber \\
&\simeq\sqrt{\frac{2}{\pi}}\left(\frac{mc^2}{\kB T R}\right)^{1/2}\times\frac{mc^2}{\kB T R^2}\left[1+{\cal O}\left(R^{-1}\right)\right]. 
\end{align}
If $R\gg mc^2/\kB T$ holds, this probability is much less than that of the test particle initially moving inward falling into the black hole.

\section{Electric current on the surface of a clump}\label{Current-2}

\begin{figure}[!h]
\centering\includegraphics[width=11cm]{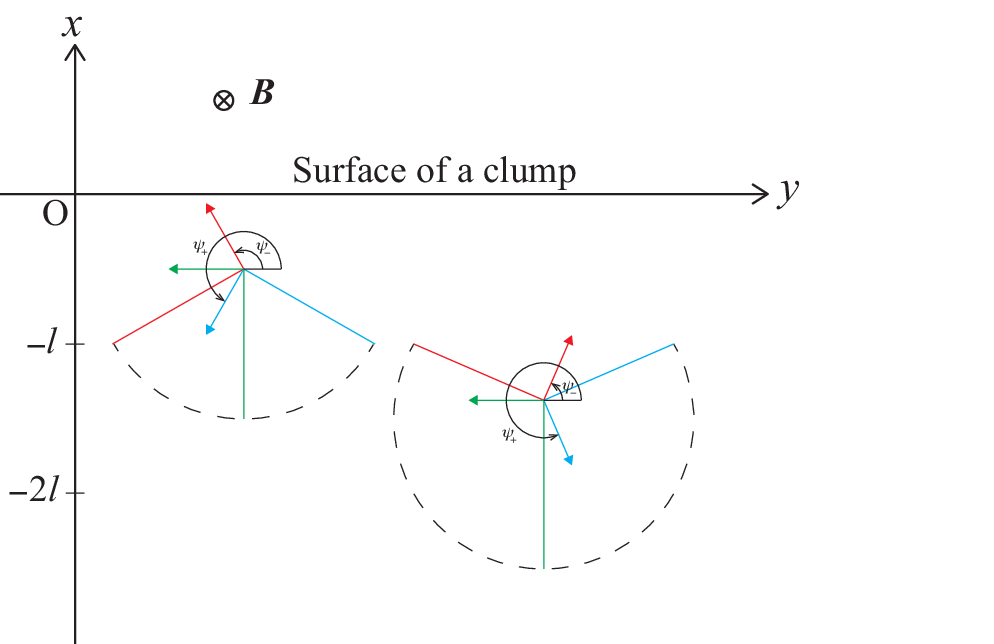}
\caption{ A neighborhood of the surface at $x=0$ of a clump which occupies the domain $x<0$ is depicted. The centers of the circular orbits 
of particles are located in the  domain $x\leq -l$. The velocities of particles located at $x$ are isotropic in the case of $x\leq -2l$, whereas those are anisotropic 
in the case of $-2l<x\leq0$.  
}
\label{fig:Surface-current}
\end{figure}

In the case that charged test particles distribute homogeneously in a clump considered in Sec.~\ref{Current-1}, 
the electric current exist only on the surface of the clump. 
We focus on the situation where the electric charge $q$ of test particles are identical to each other and is positive, and 
hence the orbital radii of the charged particles are identical to $l:=r\vartheta$. Then we consider the situation depicted in 
Fig.~\ref{fig:Surface-current}. 

The velocities of particles are represented in the form
\begin{align}
v_{(x)}&=-v\sin\psi,  \\
v_{(y)}&=-v\cos\psi,
\end{align}
where 
\begin{equation}
v:=\sqrt{v_{(\theta)}^2+v_{(\varphi)}^2}=\frac{c\calK}{R}
\end{equation}
[see Eq.~\eqref{v-L-relation}].  

Assume that the particles are uniformly distributed with the number density $n = n_0$ is constant 
only within the region $x \leq 0$ in their initial state, and that the velocity distribution is isotropic, i.e., 
the probability with respect to $\psi$ is uniform. 
Consequently, particles in the domain $-2l< x \leq0$ that satisfy $0\leq\psi<\psi_{-}$ and $\psi_{+}<\psi<2\pi$ oscillate 
between the domains $x>0$ and $x\leq0$, where 
\begin{align} 
\psi_-&=\frac{1}{2}\pi+\arcsin\left(\frac{x+l}{l}\right), \\
\psi_+&=\frac{3}{2}\pi-\arcsin\left(\frac{x+l}{l}\right) 
\end{align}
(see Fig.~\ref{fig:Surface-current}). Thus, the number density of the particles in the domain $-2l<x<2l$ 
temporally varies, whereas it in the domain $x \leq -2l$ remain constant at all times.
Furthermore, assume that all particles entering the region $x > 0$ have been removed for some reason. 
Then, remaining particles in the domain $-2l<x\leq0$ will have the velocity distribution that is uniform 
in the domain $\psi_-\leq \psi\leq\psi_+$ but vanishes outside it, and their number density is given as 
\begin{equation}
n=n_0\times\frac{\psi_{+}-\psi_{-}}{2\pi},
\end{equation}
The resultant configuration is stationary, and the expectation value of 
the velocity of each particle in $-2l\leq x\leq 0$ is given as
\begin{align}
\langle v_{(x)}\rangle &=-\frac{v}{\psi_{+}-\psi_{-}}\int_{\psi_-}^{\psi_+} \sin\psi d\psi=0,  \\
\langle v_{(y)}\rangle &=-\frac{v}{\psi_{+}-\psi_{-}}\int_{\psi_-}^{\psi_+} \cos\psi d\psi 
=\frac{2 v}{\psi_+ -\psi_-}\cos\left[\arcsin\left(\frac{x+l}{l}\right)\right].
\end{align}
Then the electric current for $(r,r+dr)$ is equal to 
\begin{align}
dI
&=q dr \int_{-2l}^0 n \langle v_{(y)}\rangle dx \nonumber \\
&=\frac{1}{\pi}qn_0 vdr \int_{-2l}^0 \cos\left[\arcsin\left(\frac{x+l}{l}\right)\right]dx \nonumber \\
&=\frac{1}{2}qn_0vldr \nonumber\\
&=\frac{1}{2}qn_0vr\sin\vartheta dr.
\end{align}

From Eq.~\eqref{theta-ell-QM}, we have 
\begin{equation}
vr\sin\vartheta=\frac{c\rg\calK^2}{\sqrt{\calQM^2+\calK^2}}.
\end{equation}
Because of $\calQM^2\gg\calK^2$ in the situation of our interest, 
we have $vr\sin\vartheta \propto m$ [see Eq.~\eqref{calQM}], 
and hence the contribution of protons to the surface electric current 
is approximately 2000 times greater than that of electrons 
in the case of the plasma clump composed of protons and electrons.

\end{document}